\documentclass[pra,bibnotes,twocolumn]{revtex4}
\usepackage{graphicx}
\begin{document}
\draft
\def\ds{\displaystyle}
\title{ Non-Hermitian anomalous skin effect }
\author{C. Yuce }
\address{Department of Physics, Eskisehir Technical University, Eskisehir, Turkey }
\email{cyuce@eskisehir.edu.tr}
\date{\today}
\begin{abstract}
A non-Hermitian topological insulator is fundamentally different from conventional topological insulators. The non-Hermitian skin effect arises in a nonreciprocal tight binding lattice with open edges. In this case, not only topological states but also bulk states are localized around the edges of the nonreciprocal  system.  We discuss that controllable switching from topological edge states into topological extended states in a chiral symmetric non-Hermitian system is possible. We show that the skin depth decreases with non-reciprocity for bulk states but increases with it for topological zero energy states.
\end{abstract}
\maketitle

The non-Hermitian extensions of topological insulators and superconductors have recently attracted great deal of attention \cite{sondeney1,1d5,ghatakdas,cyek1,cyek2,cyek3,cyek4,cyek5,cyek6}. Topological states in various systems such as complex extension of the Su-Schrieffer-Heeger (SSH) model \cite{1d1,1d2,1d3,1d3ekl,1d6,1d7,1d8,1d9,1d10,1d11,1d12,1d13,1d14,1d15,1d16,floquet1,floquet2,bhjkl,feng}, 
Aubry-Andre chain with gain and loss \cite{aah1} and complex Kitaev model \cite{kita1,kita2,kita3,kita4,kita5} have so far been explored. It was shown that topological edge states with real eigenvalues can appear in some non-Hermitian settings \cite{aah1}. Topological edge states that have complex energy eigenvalues can be used as a topological laser \cite{laser} or spontaneous topological pump at large times \cite{yucepump}. Non-Hermiticity arises from onsite gain and loss and/or nonreciprocal (asymmetrical) hopping amplitudes in a lattice. Initial attempts explored the effects of non-Hermiticiy on topological edge states already present in Hermitian systems. Takata and Notomi recently showed that topological phase transition can even be induced solely by gain and loss \cite{takata}. \\
Topological gapless edge states in a system with open boundary condition (OBC) are predicted using topological invariants, which can be computed for periodical boundary condition (PBC). This is the essence of the principle of bulk-boundary correspondence. Unfortunately the standart bulk-boundary correspondence does not always work in non-Hermitian topological systems. Furthermore, spectra in some non-Hermitian systems depend sensitively on boundary conditions in sharp contrast with topological Hermitian systems. A nonreciprocal system under PBC has an effective imaginary magnetic flux, which makes the spectrum fully complex. However, the system under OBC has not such an effective imaginary magnetic flux and the PBC and OBC predict different topological phase transition points. The extension of topological numbers to non-Hermitian systems is not straightforward, either \cite{winding1,winding2,winding3}. The so-called non-Hermitian skin effect arises in a nonreciprocal tight binding lattice with open edges \cite{bulkboun01,bulkboun02,bulkboun02b,bulkboun03,bulkboun04,bulkboun04b,bulkboun06,bulkboun07,bulkboun08,bulkboun09,bulkboun10,bulkboun11,bulkboun12,bulkboun13,bulkboun14}. In this case, not only topological states but also bulk states are localized around either edge of the nonreciprocal  system. This can be understood as an amplification of the eigenstates in one way and a corresponding decaying in the opposite way due to an imaginary
gauge field \cite{bulkboun07}. We note that skin modes are not topological so they are not immune to disorder. This topic is under hot discussion and new ideas such as the existence of hybrid skin-topological modes in a 2-dimensional system has been predicted \cite{skin01}. The bulk-boundary correspondence in non-Hermitian systems has not yet been fully understood.\\
Non-Hermitian anomalous skin effect states that bulk states in addition to topological edge states are localized around the edges in a nonreciprocal lattice. In this Letter, we show that there exists some nonreciprocal lattices whose bulk states shift towards edges but topological edge states become extended \cite{robustbulks}. This leads to the breakdown of the conventional bulk-boundary correspondence, which states that topological edge states are localized around the interface where topological phase transition occurs. We provide a simple mathematical explanation why delocalized topological zero energy states appear in non-Hermitian systems.\\
{\textbf{Nonreciprocal lattice}}: Consider a generic tight binding non-reciprocal lattice with asymmetric forward and backward hopping amplitudes. The corresponding Hamiltonian reads
\begin{eqnarray}\label{kiRYazm}
H=\sum_{n=1}^{N-1} t_n~{c^\dagger_{n+1}} c_{n} + t_n^{\prime}~{c^\dagger_{n}} c_{n+1}   
\end{eqnarray}
where $t_n$ and $t_n^{\prime}$  are site-dependent forward and backward hopping amplitudes, respectively and $\ds{{c}_{n}}$ and $\ds{{c}^\dagger_{n}}$ are the annihilation and creation operators localized at the lattice site $n$, respectively and $N$ is the total number of lattice sites. The Hamiltonian is Hermitian iff $t_n^{\star}={t_n}^{\prime}$ for all $n$. If there are $m$ lattice sites in a unit cell, then we require $t_{n+m}=t_n$ and $t_{n+m}^{\prime}=t_{n}^{\prime}$. We depict our system for $m=2$ and $m=3$ cases in Fig.1. This simple Hamiltonian allows us to explore unexpected topological features of non-reciprocal lattice with open boundary conditions.\\
As a special case, consider now a non-reciprocal lattice with alternating hopping amplitudes where $\ds{t_{n+2}=t_n}$ and $\ds{t_{n+2}^{\prime}=t_n^{\prime}}$ as shown in Fig. 1 (a). Open boundaries break translational invariance and a simple analytical formula is generally not available. Fortunately, the non-Hermitian Hamiltonian under the periodic boundary condition can be written as
\begin{eqnarray}\label{kiRYazm}
\mathcal{H} (k)=\left(\begin{array}{cc} 0& t_1+{t_2}~e^{ik} \\    t_1^{\prime}+t_2^{\prime}~e^{-ik}  & 0 \end{array}\right)
\end{eqnarray}
The corresponding right eigenvector is given by $|\psi_{\mp}^R(k)>=\frac{1}{\sqrt{2}} \left(\begin{array}{cc}\mp \sqrt{   \frac{t_1~+~t_2~e^{ik} }{  ~t_1^{\prime}~+~t_2^{\prime}~e^{-ik} }      }  &,1 \end{array}\right)^T$ and the left eigenvector reads $<\psi_{\mp}^L(k)|=\frac{1}{\sqrt{2}} \left(\begin{array}{cc}\mp \sqrt{   \frac{ ~t_1^{\prime}~+~t_2^{\prime}~e^{-ik}   }{      t_1~+~t_2~e^{ik}  }       }  &,1 \end{array}\right)$. One can easily see that this Hamiltonian has chiral symmetry: $\ds{\sigma_z\mathcal{H} (k) \sigma_z=-\mathcal{H} (k) }$ where $\sigma_i$ refers to Pauli matrices. This implies that eigenvalues  come in pairs at a given $k$. They are given by $E_{\mp}=\mp\sqrt{( t_1+{t_2}~e^{ik})  (   t_1^{\prime}+t_2^{\prime}~e^{-ik}  )   }$. Two exceptional points occur at $k=\mp\pi$ when $t_1=t_2$ and $t_1^{\prime}=t_2^{\prime}$.
\begin{figure}[t]\label{290üAik0}
\includegraphics[width=5.3cm]{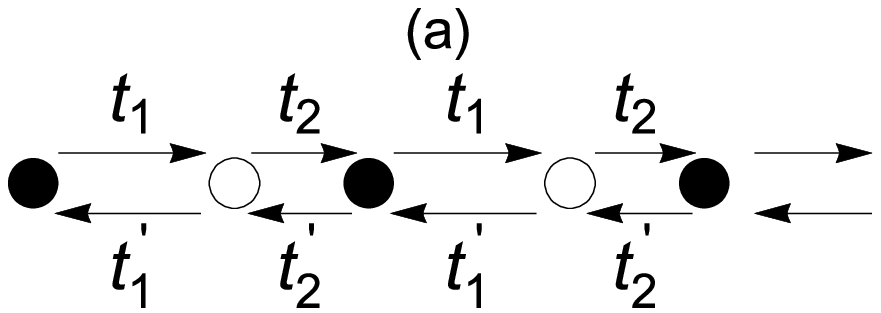}
\includegraphics[width=5.3cm]{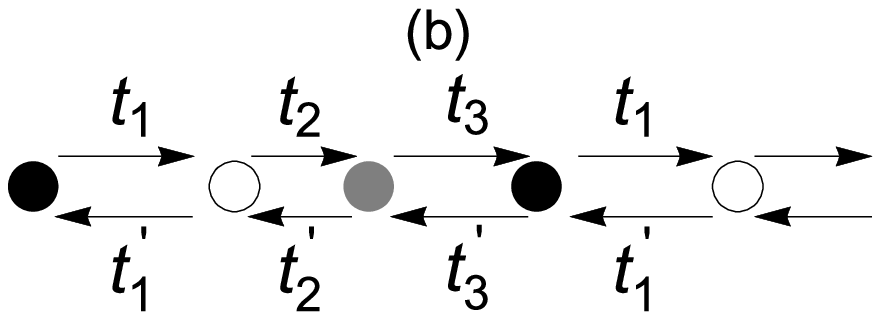}
\caption{ Non-reciprocal lattice with asymmetric forward and backward hopping amplitudes when there are $2$ (a) and $3$ (b) lattice sites in a unit cell. }
\end{figure}\\
Our aim is to study topological features of this Hamiltonian. In Hermitian systems, topological numbers are defined to predict topological phase transition point. In 1D topological systems, the winding number can be used to study Chern insulators. It takes an integer value in a topologically nontrivial system while it is zero for a trivial one. In non-Hermitian systems, the standart formula for the winding number does not work. Therefore, new topological invariants have been introduced in the literature. They are the complex winding number $\nu_{\mp}$ and the winding number of energy $\nu_{E}$ \cite{winding1}
\begin{eqnarray}\label{0yturjduc}
\nu_{\mp}=\frac{1}{\pi}\int dk    <\psi_{\mp}^L |i\partial_k|  \psi_{\mp}^R>   \nonumber\\
\nu_{E}=\frac{1}{2\pi}\oint dk ~\partial_k    Arg(E_+-E_-) 
\end{eqnarray}
\begin{figure}[t]\label{2678ik0}
\includegraphics[width=4.25cm]{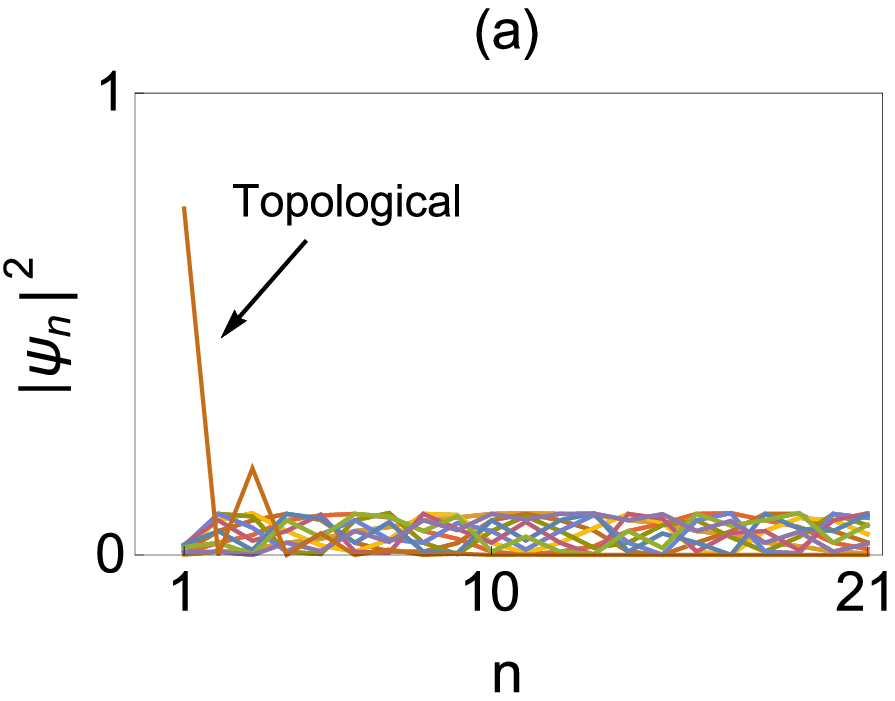}
\includegraphics[width=4.25cm]{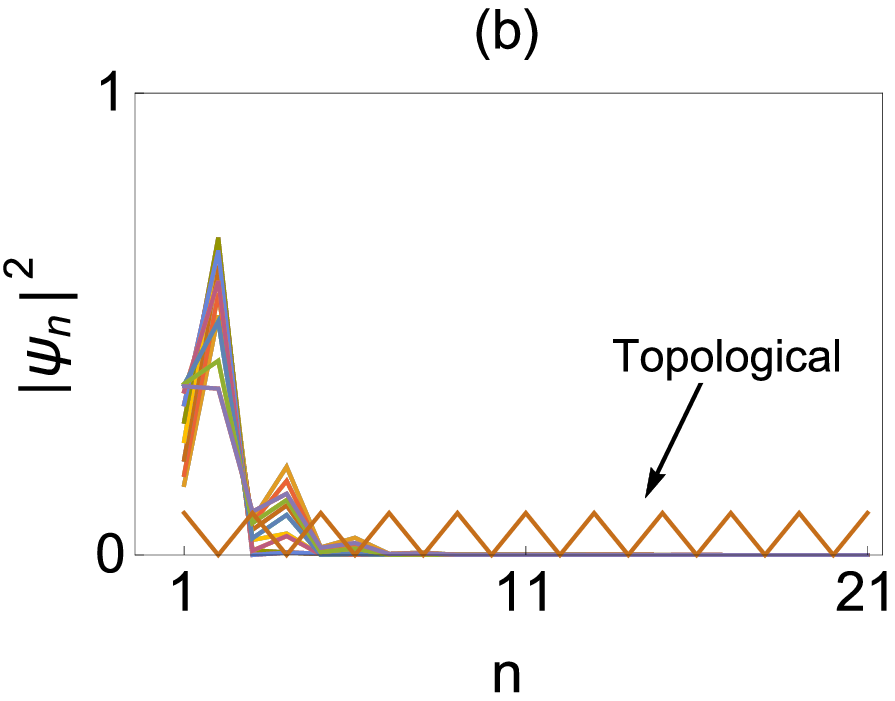}
\includegraphics[width=4.25cm]{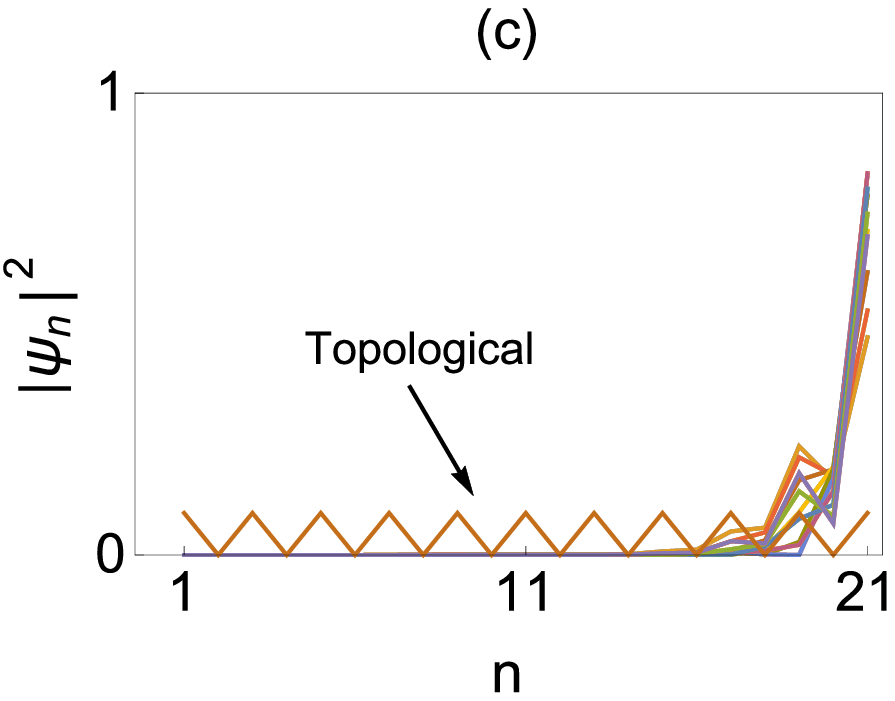}
\includegraphics[width=4.25cm]{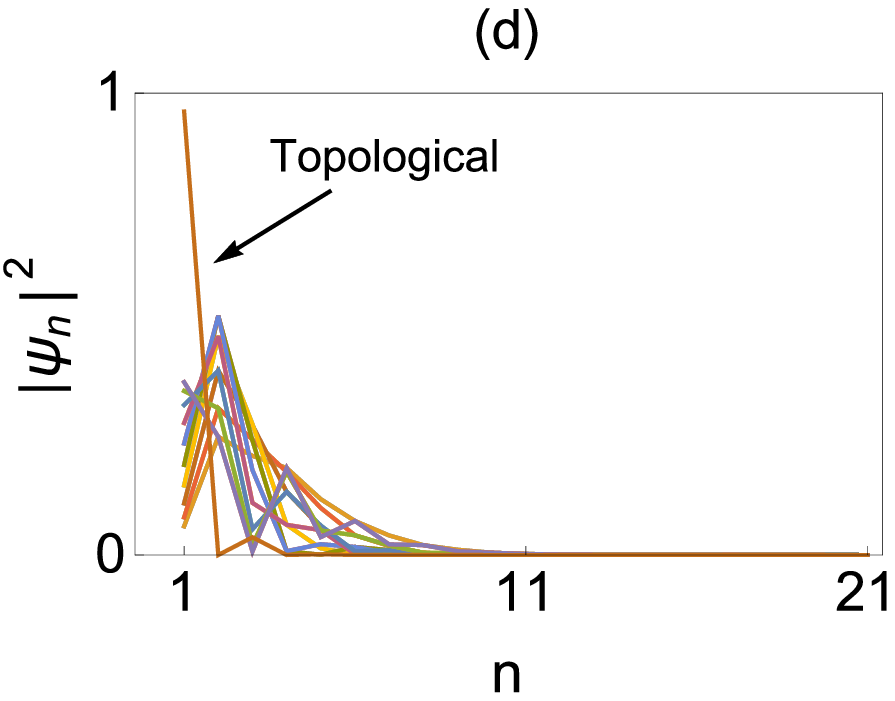}
\includegraphics[width=4.25cm]{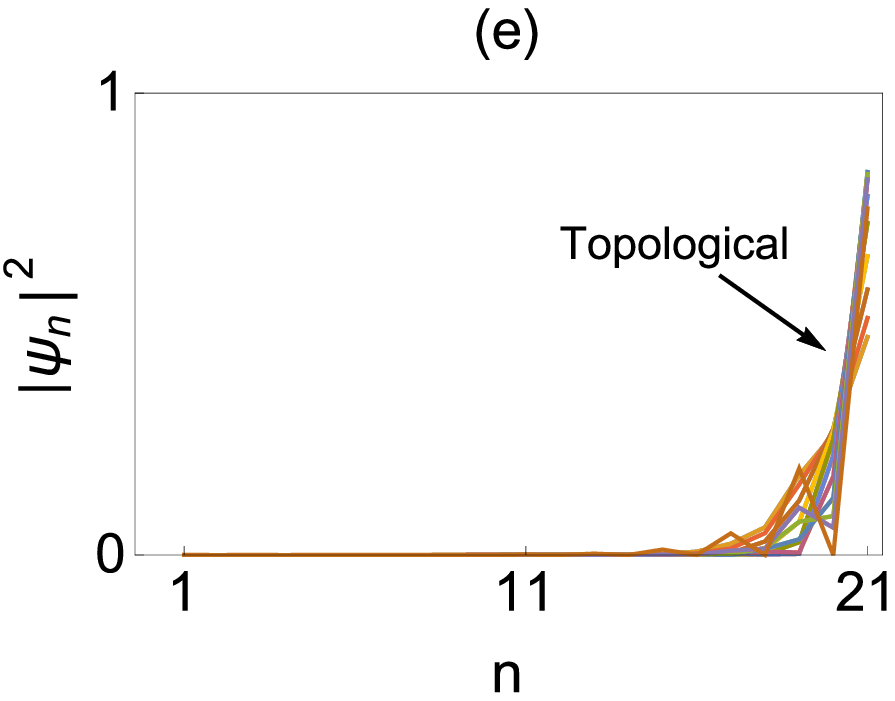}
\includegraphics[width=4.25cm]{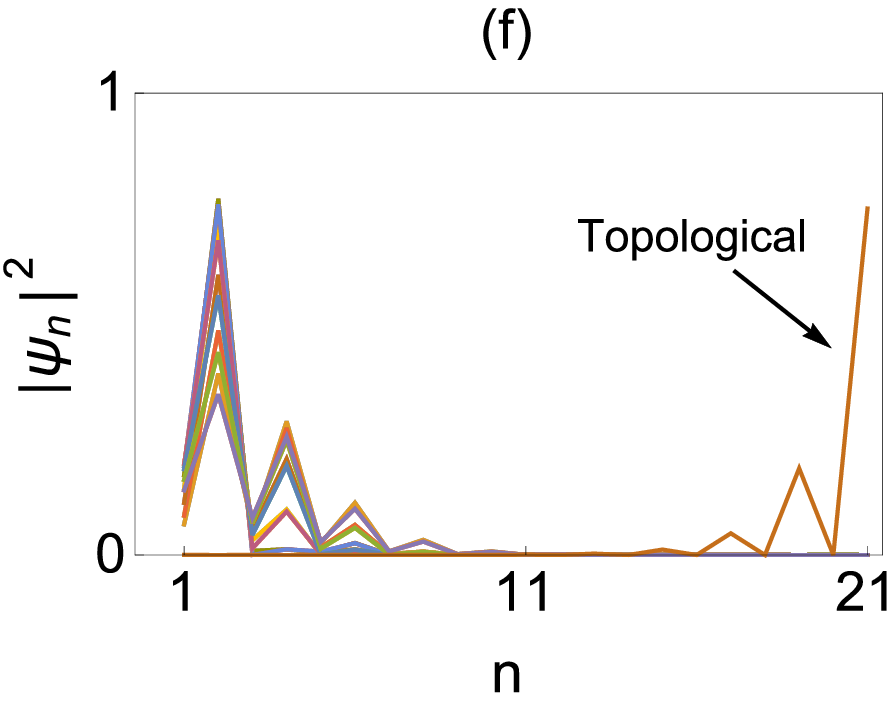}
\caption{ The density profiles of all eigenstates for a non-reciprocal lattice with alternating hopping amplitudes when $N=21$. $(t_1,t_2)=(0.5,1)$ are fixed for all plots. $(t_1^{\prime},t_2^{\prime})$ parameters are $(0.5,1)$ (a), $(1,0.05)$ (b), $(1,5)$ (c), $(0.2,0.5)$ (d) and $(2,4)$ (e) and $(2,0.05)$ (f). Due to the non-Hermitian anomalous skin effect, topological zero energy edge state becomes extended along the lattice while the bulk states are shifted towards one edge as seen in (b) and (c). Due to the non-Hermitian skin effect, both topological zero energy edge state and bulk states are shifted towards either the same or opposite edges as seen in (d), (e) and (f).  }
\end{figure}
where $\ds{|  \psi_{\mp}^R>  }$, $\ds{|  \psi_{\mp}^L>  }$ are the normalized right and left eigenvectors of the Hamiltonian, and the integral is taken over the 1D Brillouin zone (BZ). It was shown in \cite{winding1} that the complex winding number is half of the summation of the two winding numbers of the real part of Hamiltonian surrounding the EP point when k ranges from $-\pi$ to $\pi$ in the space spanned by the real part of $h_x$ and $h_y$, where $\mathcal{H} (k)=h_x\sigma_x+h_y\sigma_y$. Therefore, it is $1/2$ if an EP is enclosed or $1$ if two EPs are enclosed. If no EP is enclosed, it is zero. The winding number of energy is $0.5$ when $t_1<t_2$ and $ t_1^{\prime}>t_2^{\prime}~$ and $0$ otherwise.\\
Consider now that the system has open edges. Since the conventional bulk-boundary correspondence is broken in non-Hermitian systems, the behaviors under PBC and OBC are quite different. In the Hermitian case, the hopping parameters plays a vital role for the existence of topological zero energy states. If $t_1<t_2$, then topological zero energy states appear. Because of the non-reciprocity of the hopping amplitudes, edge state and bulk states are expected to be shifted towards one edge due to the non-Hermitian skin effect. We find that the parity of total number of lattice sites $N$ is of great importance in our system. Below we will show that topological zero energy states always exist for any values of hopping parameters when $N$ is an odd number. But this is not the case when $N$ is an even number, where topological zero energy states appear only for some certain relations between the hopping amplitudes are satisfied. Let us now study topological zero energy states when $N$ is an odd number. In Fig 2, we plot the absolute squares of all eigenstates for various values of backward hopping parameters $t_1^{\prime}$ and $t_2$ at fixed $t_1=0.5$ and $t_2=1$ when $N=21$. The Fig 2 (a) is for the Hermitian system and the topological edge state occurs at the edge and all of the bulk states are extended all over the lattice as expected. The topological edge state are localized at the left edge since the topological phase transition occurs between the vacuum and the nontrivial left edge with $t_1<t_2$. We now change $t_1^{\prime}$ and $t_2^{\prime}$ values at fixed $t_1$ and $ t_2$ to see the effect of the non-reciprocity. Surprisingly, we find that the topological edge state becomes extended while all of the bulk states are shifted towards one edge as can be seen from Fig 2 (b) and (c). The localization of the bulk states around one edge can be understood using the non-Hermitian skin effect. However, the existence of the extended topological zero energy states is unexpected. This is the main finding of this paper and we introduce non-Hermitian anomalous skin effect. In the non-Hermitian skin effect, the skin depth decreases with increasing non-reciprocity. Conversely, the skin depth increases with $t_1^{\prime}$ at fixed $t_1$ and $t_2$ and becomes comparable to the system size at certain values of $t_1^{\prime}$ in the non-Hermitian anomalous skin effect. According to the standard bulk-boundary correspondence, topological states occur around the edges where topological phase transition occurs. This standart view is broken in our system. In Fig 2 (d-f), we see that all states (topological+bulk) are localized around the edges due to the non-Hermitian skin effect. We emphasize that  they are localized around the same edge in (d) and (e) while the edge state is separated from the bulk state in (f). The case in (f) is of importance since all of the bulk states are well separated spatially from the topological zero energy state.\\
The most interesting feature of the topological edge states is that they are robust against certain types of disorder. In our system, the disorder are required not break the chiral symmetry. We analyze robustness of the edge states in our system against hopping amplitude disorder by introducing randomized hopping amplitudes in the lattice, which maintains the chiral symmetry. In our numerical computation, we introduce randomized coupling all over the lattice as $t_n{\rightarrow}t_n+\delta_n$ and $t_n^{\prime}{\rightarrow} t_n^{\prime}+\delta_n^{\prime}$, where $\delta_n^{\prime}$ and $\delta_n^{\prime}$ are real-valued random set of constants. Therefore, the hopping amplitudes between the neighboring sites become completely independent in both forward and backward directions. We find that these edge states resist the disorder, i. e., their eigenvalues are always equal to zero. We see that this is true even if $\delta_n>t_n$ and $\delta_n^{\prime}>t_n^{\prime}$. This is expected because of the topological nature of the edge states. However, the energy eigenvalues for the bulk states change considerably with the disorder. This shows us that topological edge states are immune to the disorder.
\begin{figure}[t]\label{2678ik0}
\includegraphics[width=4.25cm]{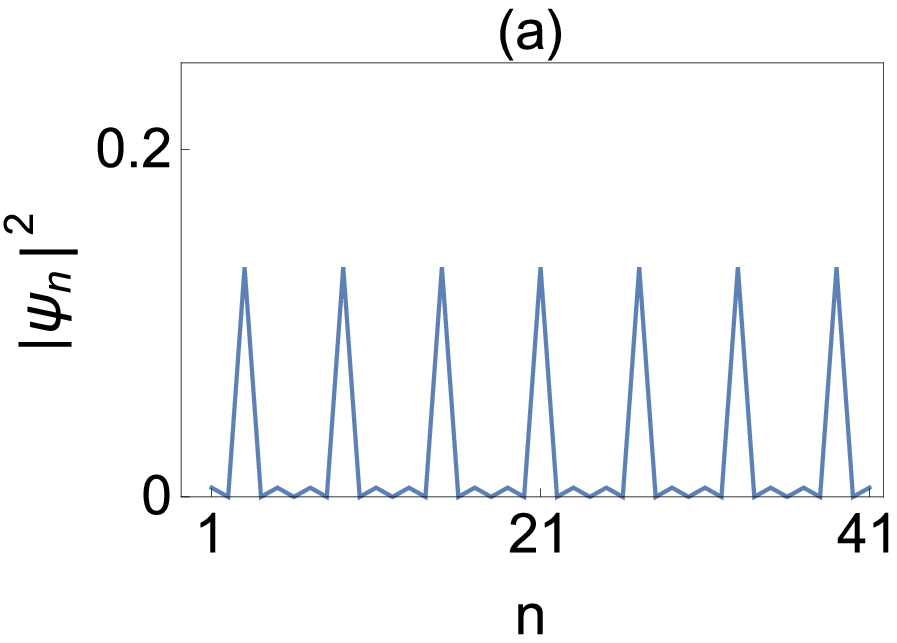}
\includegraphics[width=4.25cm]{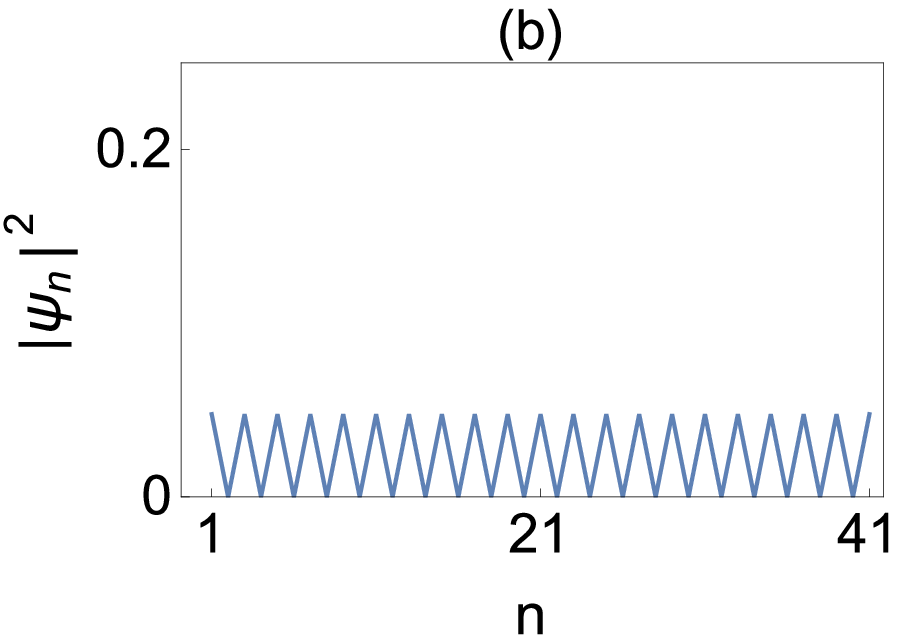}
\caption{The density profiles of topological zero energy states for a non-reciprocal three-band (a) and four-band (b) systems with $N=41$. The parameters are given by $(t_1,t_2,t_3)=(1,0.2,1)$, $(t_1^{\prime},t_2^{\prime},t_3^{\prime})=(1,1,0.2)$ (a) and $(t_1,t_2,t_3,t_4)=(1,0.5,1,2)$, $(t_1^{\prime},t_2^{\prime},t_3^{\prime},t_4^{\prime})=(0.5,0.5,2,1)$ (b). The topological zero energy states are extended, which implies that the standard bulk-boundary correspondence is broken. }
\end{figure}\\
Let us analyze the appearance of topological zero energy modes and their robustness. We have shown that topological zero energy states exist if there are two lattice sites in a unit cell. A question arises. Do topological zero energy states exist if there are an arbitrary number of lattice sites in a unit cell? The answer is Yes. To see it, we note that the Hamiltonian (\ref{kiRYazm}) has chiral symmetry. The chiral operator is given by $\ds{C=diag(1,-1,1,-1,...,1,-1,1)}$, which satisfies $\ds{ CHC^{-1}=-H }$ where $H$ is the matrix form of the Hamiltonian with open edges. The chiral symmetry remains intact even for the disordered Hamiltonian. For a chiral symmetric Hamiltonian, the energy eigenvalues come in pairs $(E,-E)$. If the number of lattice sites $N$ is an odd number, then one of the states is not paired. Consequently, there exists a state with zero energy eigenvalue, regardless of the number of lattice sites in a unit cell as long as $N$ is an odd number. But zero energy eigenstate don't necessarily appear when $N$ is an even number. The zero energy state is robust against the hopping amplitude disorder since the chiral symmetry remains intact, which implies that zero energy state is still not paired. To check our discussion, we consider three-band $t_{n+3}=t_n$ and $t_{n+3}^{\prime}=t_{n}^{\prime}$ and four-band systems $t_{n+4}=t_n$ and $t_{n+4}^{\prime}=t_{n}^{\prime}$. In Fig 3, we plot the density profile for the topological zero energy states for a three-band (a) and a four-band systems at $N=41$. The topological zero energy states appear in both cases and are robust against the hopping amplitude disorder. As can be seen, they are extended along the lattice. Note that bulk states move to one side as a result of the non-reciprocity. \\
The chiral symmetry plays a role on the existence of the extended topological zero energy states. Let $\psi_{E}(n)$ be the complex amplitude at the lattice site $n$ corresponding to the state with energy $E$. Due to the chiral symmetry,  $\psi_{-E}(n)=C \psi_{E}(n)$, where $C$ is the chiral operator introduced above. Topological zero energy modes are their own chiral-symmetric partners and hence $\psi_{0}(n)=C \psi_{0}(n)$. This implies that $\psi_{0}(n)$ vanishes for all even number of $n$ ($\psi_{0}(n)=0$ when $n=2,4,...,N-1$). The non-Hermitian skin effect forces the density profiles of the eigenstates to move in one direction but the chiral symmetry makes an extra requirement on the topological zero energy eigenstate. These in turn lead to the appearance of the extended zero energy state. As $t_1^{\prime}$ increases, then zero energy eigenstate move towards the other edge and extends to lattice sites at the other edge. The zero energy edge state becomes localized around the other edge when $t_1^{\prime}>t_2$.\\
Non-Hermiticity can arise not only from non-reciprocity but also from gain/loss. We extend our formalism to the non-reciprocal system with alternating gain and loss. Consider the non-reciprocal lattice with alternating gain and loss 
\begin{eqnarray}\label{kiRYazDJSALm}
H_{\gamma}=H  + \sum_{n=1}^{N} i  ~\gamma ~ (-1)^n~{c^\dagger_{n}} c_{n}
\end{eqnarray}
where $H$ is given in (\ref{kiRYazm}) and $\gamma$ is the non-Hermitian degree and $N$ is an odd number. For example, if there are two sites in a unit cell, then the above Hamiltonian takes the form of $\mathcal{H}_{\gamma}(k)=\mathcal{H} (k)+i\gamma~\sigma_z$, where $\mathcal{H} (k)$ was given in (\ref{kiRYazm}).\\
We stress that either complex winding number or winding number of energy (\ref{0yturjduc}) can't predict topological modes for the Hamiltonian (\ref{kiRYazDJSALm}). To explore topological feature of this Hamiltonian, we start with the shifted Hamiltonian $\ds{H_{\gamma}^{\prime}=H_{\gamma}+  \sum_{n=1}^{N} i  ~\gamma ~{c^\dagger_{n}} c_{n} }$. In this case, we get a system that has gain (loss) at every even number of lattice sites when ${\gamma}>0$ (${\gamma}<0$). Since we just shift the energy eigenvalues, the eigenstates of $H_{\gamma}$ and $H_{\gamma}^{\prime}$ are the same. Furthermore, $\psi_{0}(n)$ (zero energy eigenstate of $H$) is simultaneous eigenstate of both $H$ and $H_{\gamma}^{\prime}$. This is because of the fact that $\psi_{0}(n)$ vanishes at n=2,4,...,N-1, where gain/loss are introduced. Therefore we conclude that $\psi_{0}(n)$ is also an eigenstate of $H_{\gamma}$, with eigenvalue $-i~{\gamma}$. In other words, the form of the topological state are the same with or without alternating gain and loss. But they have zero and non-zero energy eigenvalues. This is true even in the presence of chiral symmetry protecting disorder. Note that all bulk states change their form with gain and loss. As a result, we say that $H_{\gamma}$ has a non-zero topological state with purely energy eigenvalue $-i~{\gamma}$, which is robust against the hopping amplitude disorder, i. e., the eigenvalue $-i~{\gamma}$ remain the same under such a disorder. This was called pseudo topological insulator \cite{pseudo}. \\
To sum up, we have explored topological zero energy modes protected by chiral symmetry in a nonreciprocal tight binding lattice with an odd number of lattice sites. We have proposed the idea of non-Hermitian anomalous skin effect. In a non-reciprocal lattice lattice under open boundary conditions, bulk states are shifted towards one edge of the lattice. In this case, the skin depth decreases with increasing non-reciprocity. Conversely, the skin depth increases with increasing non-reciprocity for topological zero energy states if the total number of the lattice sites is an odd number.  In this way, controllable switching from topological edge states into topological extended states in a non-Hermitian system is possible. This leads to the breakdown of the famous concept in topological insulators, which states that topological states are localized around the interface where topological phase transition occurs. We have also considered additional gain and loss in the nonreciprocal lattice and shown that introducing alternating gain or loss in to the system has no influence on the form of topological modes.

\end{document}